\documentclass{emulateapj} 

\shorttitle{V5116 Sgr, an eclisping supersoft post-outburst nova?}
\shortauthors{Sala, Hernanz, and Ferri}

\begin{document}

\title{V5116 Sgr, an eclipsing supersoft post-outburst nova?}

\author{G. Sala\altaffilmark{1}, 
	M. Hernanz\altaffilmark{2},
 	C. Ferri\altaffilmark{2},
	J. Greiner\altaffilmark{1}}

\altaffiltext{1}{Max-Planck-Institut f\"ur extraterrestrische Physik,
PO Box 1312, D-85741 Garching b.M., Germany; gsala@mpe.mpg.de, jcg@mpe.mpg.de}
\altaffiltext{2}{Institut de Ci\`encies de l'Espai (CSIC-IEEC), 
Campus UAB, F. Ci\`encies, 
08193 Bellaterra, Spain; hernanz@ieec.uab.es, ferri@ieec.uab.es} 

\begin{abstract}
V5116 Sgr (Nova Sgr 2005 No. 2), discovered on 2005 July 4, 
was observed with XMM-Newton in March 2007, 20 months after the optical outburst. 
The X-ray spectrum shows that the nova had evolved to a pure supersoft X-ray source, 
with no significant emission at energies above 1~keV.
The X-ray light-curve shows abrupt decreases and increases 
of the flux by a factor $\sim8$. 
It is consistent with a periodicity of 2.97~h,
the orbital period suggested by \cite{dob07}, 
although the observation lasted just a little more than a whole period.
We estimate the distance to V5116~Sgr to be $11\pm3\,kpc$.
A simple blackbody model does not fit correctly the EPIC spectra, with  
$\chi^2_{\nu}>4$. In contrast, ONe rich white dwarf atmosphere models provide 
a good fit, with 
$N_{\rm H}= 1.3(\pm0.1)\times 10^{21}\,cm^{-2}$, 
$T=6.1(\pm0.1)\times10^5\,K$, and 
$L=3.9(\pm0.8)\times10^{37}(D/10~\rm{kpc})^2{\rm erg}\,{\rm s}^{-1}$
(during the high-flux periods). 
This is consistent with residual hydrogen burning in the white dwarf envelope. 
The white dwarf atmosphere temperature is the same both in the low and the high 
flux periods, ruling out an intrinsic variation of the X-ray source as the origin of the flux changes.
We speculate that the X-ray light-curve may result from a partial coverage by an asymmetric 
accretion disk in a high inclination system.

\end{abstract}


\section{Introduction}

Classical novae occur in close binary systems of the cataclysmic variable type, 
when a thermonuclear runaway results in explosive 
hydrogen burning on the accreting white dwarf (WD). It is theoretically predicted that 
novae return to hydrostatic equilibrium after the ejection of a fraction of the 
accreted envelope.
Soft X-ray emission arises in some post-outburst novae
as a consequence of residual hydrogen burning on top of the WD.
As the envelope mass is depleted, the photospheric radius 
decreases at constant bolometric luminosity 
(close to the Eddington value) with an increasing effective temperature. 
This leads to a hardening of the spectrum from optical through 
UV, extreme UV and finally soft X-rays, with 
the post-outburst nova emitting as a Supersoft Source (SSS) with a 
hot WD atmosphere spectrum.
The duration of this SSS is related to 
the nuclear burning timescale of the remaining H-rich envelope 
and depends, among other factors, on the WD mass \citep{TT98,SH05a,SH05b}. 

All post-outburst novae are expected to undergo SSS emitting phase, 
provided a H-rich envelope is left. 
Nevertheless, the number and duration of SSS states observed in Galactic novae is small. 
In a systematic search for the X-ray emission from novae in the ROSAT archive,
\cite{Ori01a} found only three SSS novae from a total of 39 novae observed less 
than ten years after the outburst (GQ Mus, V1974~Cyg, N~LMC~1995).
After the ROSAT era, observations by Beppo-SAX and Chandra revealed SSS X-ray emission for 
some more post-outburst novae: V382~Vel \citep{Ori02,Bur02},
V1494~Aql, with a puzzling light curve showing a short burst 
and oscillations \citep{Dra03}, and V4743~Sgr, also with a very variable light-curve
\citep{Nes03,Pet05}. 
The SSS phase of the recurrent 
nova RS~Oph in the 2006 outburst was monitored with Swift, XMM-Newton and Chandra,
and the end of the SSS state occurred less than
100 days after outburst \citep{Bode06,Hachi07}.
XMM-Newton has contributed with some more observations of novae.
Nova~LMC 1995 showed the SSS emission 5 years after outburst \citep{Ori03}. 
Other post-outburst novae were detected with a harder spectrum 
associated to the expanding ejecta, as 
Nova LMC 2000 \citep{Gre03} and V4633 Sgr \citep{HS07}, 
or reestablished accretion flow \cite[V2487~Oph,][]{HS02,fer07}.
More recently, the SSS XMMSL1~J070542.7-381442 
was identified as a nova \citep{rea07a,rea07b,tor07}.

X-ray observations of the central area of M31,
with its moderate absorption, offer a good chance to monitor the SSS phase of novae: 
\cite{pie05} reported 21 X-ray counterparts for novae in M31 
-- mostly identified as SSS by their hardness 
ratios -- and two in M33. Following that work, XMM-Newton and Chandra
monitoring of M31 between July 2004 and February 2005 provided the 
detection of eleven out of 34 novae within a year after optical outburst \citep{pie07}.
This suggests that the low fraction of SSS found by \cite{Ori01a} was due to selection effects 
and/or too poor sampling.

X-ray observations of post-outburst novae 
provide crucial information:
soft X-rays yield a unique insight 
into the remaining burning envelope on top of the WD,
while hard X-rays reveal shocks in the nova ejecta. 
The properties of  ``quiescent novae'', once they have turned-off and once 
accretion is reestablished (which can occur before or after the turn-off), 
are revealed both in hard and soft X-rays; 
they emit then as ``standard'' cataclysmic variables.
In view of the scarcity of objects observed and the diversity of 
behaviors detected, only the monitoring of as many novae as possible,
with large sensitivity and spectral resolution (as those offered 
by XMM-Newton and Chandra) can help to understand post-outburst novae.

V5116 Sgr (Nova Sgr 2005 No. 2) was one of the targets included in our X-ray monitoring programme
of post-outburst Galactic novae with XMM-Newton. 
It was discovered by \cite{lil05} on 2005 July 4.049~UT, 
with magnitude $\sim$8.0, rising to mag 7.2 on July 5.085. 
The expansion velocity derived from a sharp P-Cyg profile
detected in a spectrum taken on July 5.099 was $\sim$1300~km/s.
IR spectroscopy on July 15 showed emission lines with FWHM~$\sim$2200~km/s \citep{rus05}. 
Photometric observations obtained during 13 nights in the period August-October 2006 
show the optical light curve modulated with a period of $2.9712\pm0.0024$~h \citep{dob07},
which the authors interpret as the orbital period. They propose that the  
light-curve indicates a high inclination system with an irradiation effect on the secondary star. 
A first X-ray observation with Swift/XRT (0.3--10~keV) in August 2005 yielded 
a marginal detection with 1.2($\pm$1.0)$\times10^{-3}$~cts/s \citep{nes07a}. Two years later,  
on 2007 August 7, Swift/XRT showed the nova as a SSS, 
with 0.56$\pm$0.1~cts/s \citep{nes07b}. A first fit with a blackbody indicated 
$T\sim4.5\times10^5K$. More recently, a 35~ks Chandra spectrum obtained 
on 2007 August 28 was fit with a WD atmospheric 
model with N$_{\rm H}=4.3\times10^{21}$~cm$^2$ and $T=4.65\times10^5K$ \citep{NelOri07}.

Here we report on the X-ray light-curve and broad-band spectra of V5116 Sgr
as observed by XMM-Newton in March 2007, 609 days after outburst. 

\section{Observations and data analysis}

V5116 Sgr was observed with XMM-Newton \citep{Jan01} on 2007 March 5
\cite[ observation ID: 0405600201, PI: Hernanz]{sal07}. 
The exposure times were 12.7~ks for the European Photon Imaging Cameras (EPIC) 
MOS1 and MOS2 \citep{Tur01}, 8.9~ks for the EPIC-pn \citep{Str01}, 
12.9~ks for the Reflection Grating Spectrometers \cite[RGS,][]{Her01}, 
and 9.2~ks for the Optical Monitor with U filter \cite[OM,][]{Mas01}.

The observation was affected by an X-ray storm, which produced
moderate background in the EPIC cameras for most of the exposure time. 
Fortunately enough, our target was at least a factor 10 brighter than the background.
In addition, the source spectrum is very soft and little affected by solar flares.
We therefore do not exclude any time interval of our exposures, but we pay 
special attention to the background subtraction of both spectra and light-curves.
Data were reduced using the XMM-Newton Science Analysis System 
(SAS~7.1.0). Standard procedures described in the SAS documentation 
\citep{sas05,abc04} were followed. 
We applied flag filter FLAG=0, and only events with pattern smaller than 12~(MOS) and 4~(pn), 
were selected for the analysis of the EPIC data.

The high flux of the source resulted in pile-up in the three EPIC cameras (used in 
large partial window mode). We minimize the pile-up effect by extracting both spectra and
light-curves from annular regions that exclude the central, piled-up pixels. 
We determined the optimal extraction region by extracting the source spectra 
using annular regions with increasing inner 
radius until we found that the pile-up effect was minimized, keeping a 
reasonable signal to noise. For the extraction of the source light-curve we use an 
annular region with inner and outer radii 25 and 38~arcsec (for pn), 
and 10 and 80~arcsec (for MOS) respectively for the whole exposure time.
Background for the MOS cameras was extracted from an annulus region centered on the source, 
with an external radius of 140~arcsec and an internal radius of 115~arcsec. 
In the case of the pn, 
the CCD where the source was detected had no large area without source photons, so
we used two box regions in different CCDs, free of sources, for the background light-curve 
determination. 
Nevertheless, we could not extract a spectrum completely free of pile-up for the pn camera
and we consider only the MOS cameras for broad-band spectral analysis. 
We increased the statistics for the MOS spectra during the low-flux periods by 
reducing the inner radii of the source extraction regions down to 5~arcsec.
For the high-flux state we increased the inner radii to 40~arcsec to
avoid pile-up as much as possible.
We have also reduced the RGS data, not affected by pile-up, using the
standard procedures. We also find two well differentiated states, coincident
in time with the high and low rate periods of the EPIC cameras. The RGS spectra, 
while compatible with the broad band models used to fit the MOS spectra in this paper, 
show a number of absorption/emission features that require a detailed 
spectral analysis, to be presented in a separate work.

\section{Light-curve}

\begin{figure}
\bigskip
\epsscale{1.1}
\plotone{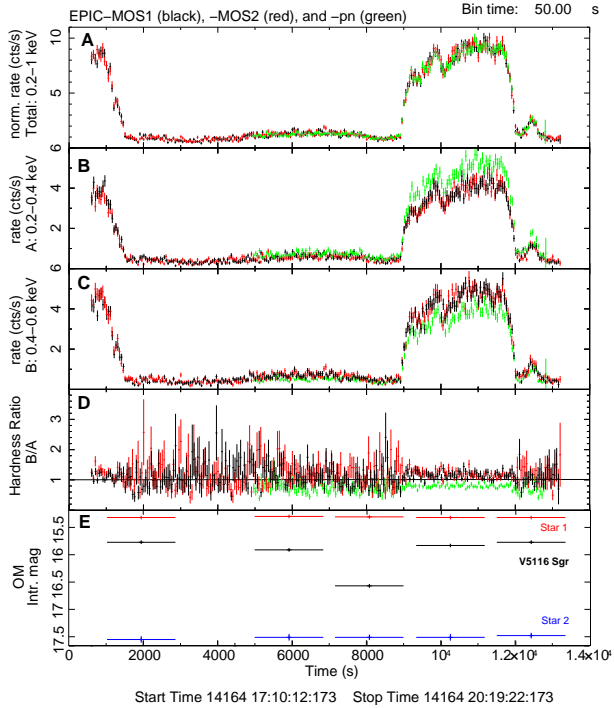}
\caption{V5116 Sgr X-ray and UV light-curves during the XMM-Newton obsevation on 2007 March 5.
Time units are TJD (JD-2440000.5) and the time bin is 50~s. 
X-ray lightcurves are background substracted.
{\bf A:} EPIC MOS1 (black), MOS2 (red), and EPIC-pn (green) count-rates, 
in the 0.2--1.0~keV energy range, normalized to 10 cts/s (see text for details).
{\bf B and C:} EPIC MOS1, MOS2 and pn count-rates in the bands 0.2--0.4~keV and 
0.4--0.6~keV. {\bf D:} Hardness ratio for the three EPIC cameras. 
{\bf E:} OM U-band instrumental magnitudes for V5116 Sgr and for two comparison stars 
in the OM FOV.}
\label{fig:lcs}
\end{figure}

Panel A in Figure~\ref{fig:lcs} shows the background subtracted light-curves 
of the three EPIC cameras in the 0.2--1.0~keV band. There is no significant flux above 1.0~keV. 
EPIC-pn experienced full scientific buffer at the beginning of the observation, and started to 
collect data later than the other instruments. The Y-axis in the EPIC light-curves is only 
in relative units, since the annulus extraction regions used do not provide
the total source rate. For comparison purposes, 
the 0.2--1.0~keV count-rates have
been renormalized for the three EPIC cameras to 10~cts/s (close to the actual 
maximum count-rate using the annulus extraction region, 
10.9~cts/s for MOS1, 10.2~cts/s for MOS2, and 13.1~cts/s in pn).

The X-ray light-curve shows clearly two states: a high-flux state, during the first 600s, and
again starting 8300~s after the start of the MOS exposures and lasting for 3000~s; and a low-flux state
between 600~s and 8300~s, and during the last 1200~s of the observation. 
A small flare is observed during the last low-flux period. 
The count-rate during the low-flux period between 600~s and 8300~s
does not show significant variations (25\% rms variability). 
In contrast, the count-rate during the high-state shows flaring activity. 
If the first decline after the initial high-flux period corresponds to the end of 
a 3-ks burst like the one starting at 8300~s, this would point to 
a periodicity in the X-ray light-curve of 10.5--10.7~ks (2.92--2.97~h).
This is consistent with the periodicity found in the optical light-curve in August--October 2006
by \cite{dob07} ($2.9712\pm0.0024$~h), which was suggested by the authors to be the orbital period.

Panels B and C in Figure~\ref{fig:lcs} show the count-rates in two energy bands, 
~0.2--0.4~keV and ~0.4--0.6~keV respectively, for the three EPIC cameras. 
EPIC-pn shows a softer spectral distribution than the two MOS cameras, 
especially during the high-flux state. This indicates that pile-up is still 
affecting the EPIC-pn data (and therefore we do not use them in the spectral analysis). 
The source is very soft with low signal above 0.6~keV, 
insufficient to define a third energy band above 0.6~keV.
The hardness ratio (Figure~\ref{fig:lcs}, panel~D) indicates that there are
no significant variations of the spectral properties during the whole observational period,
even when the rate is increasing by a factor 8--10 in the high-flux phases. 

Panel E in Figure~\ref{fig:lcs} shows the light-curve of the OM exposures in the U-band 
in instrumental magnitudes. The light-curve of two stars in the field of view close 
to V5116 Sgr are shown for comparison. The OM light-curve is 
compatible with the 2.97~h periodicity observed by \cite{dob07}, 
but its minimum is not correlated either with 
the maximum or the minimum in X-rays. This points 
to different origins for the X-ray and UV emission.

\begin{figure}
\bigskip
\epsscale{1.1}
\plotone{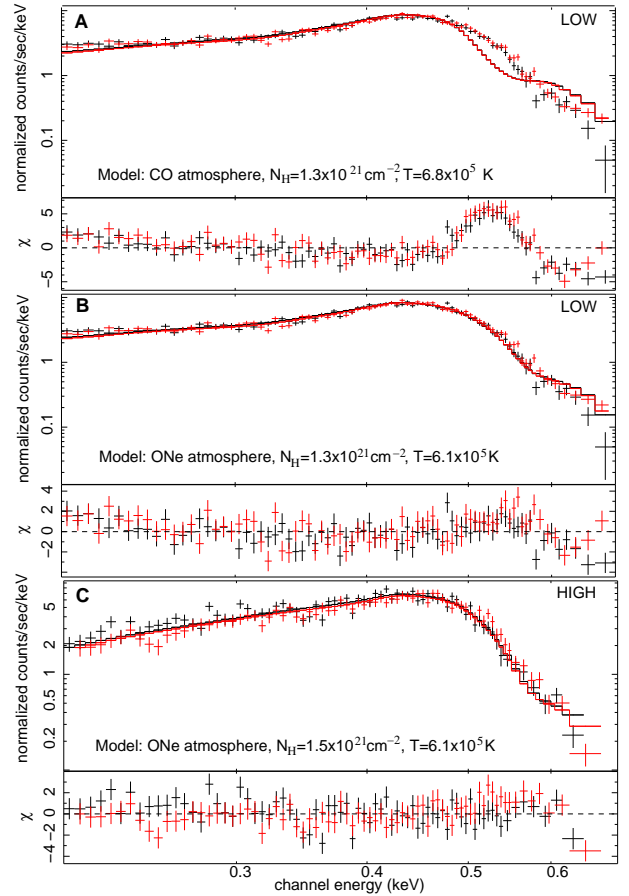}
\caption{XMM-Newton MOS1 (black) and MOS2 (red) spectra of V5116 Sgr.
{\bf A:} Low-flux spectra fit with a CO WD atmosphere model. 
{\bf B:} Low-flux spectra fit with an ONe WD atmosphere model. 
{\bf C:} High-flux MOS1 and MOS2 spectra fit with an ONe WD atmosphere model.}
\label{fig:spec}
\end{figure}

\section{Spectral analysis}

The X-Ray Spectral Fitting Package \cite[XSPEC~11.3,][]{xsp03} was used 
for spectral analysis. We used only MOS data for the spectral analysis, since
EPIC-pn is too affected by pile-up. Even for the EPIC-MOS cameras, 
some residual pile-up is likely responsible for a component at energies twice 
the maximum of the emission, i.e., 0.8-1.0~keV, caused by double events at 0.4-0.5~keV
counted as single photons with twice the energy. These lost photons are only a small 
fraction of the total at the maximum of emission (and cause a small effect in the 0.3--0.5~keV band), 
but they contribute notably at 0.8--1.0~keV, 
where the intrinsic flux of the source is very small, if any.
This implies that we can trust the spectral analysis only up to 0.7~keV and we can not
provide any constraint on a possible reestablished accretion disk.
In our spectral fits, we use the {\it phabs} model with \cite{wil00} 
abundances for the foreground absorption.

Figure~\ref{fig:spec} shows the MOS1 and MOS2 spectra during the low-flux period.  
A first attempt to fit a simple absorbed blackbody model to the data yields an 
unacceptable fit, with $\chi^2_{\nu}>4$.
Therefore we used \cite{MV91} WD atmosphere emission models, gently 
provided by Jim MacDonald, and 
built tables to be read as external models in XSPEC.
Previous studies have shown the importance of including 
WD atmosphere models instead of simple blackbodies 
for a correct broad-band spectral analysis of novae in the SSS phase \citep{Bal98,BK01,OG99}. 
Two sets of models are available, corresponding to CO and ONe compositions of 
the WD. 
As shown in Figure~\ref{fig:spec} (panel A), CO atmosphere models do not fit well the data.
The main difference between the two types of nova atmospheres at these temperatures is 
the hard energy tail of ONe models, which is suppressed in CO models 
due to the C ionization edges (CV at $\sim$0.4~keV and CVI at $\sim$0.5~keV).
The best fit is obtained with the ONe models 
(Figure~\ref{fig:spec}, panel B), with 
$N_{\rm H}=1.3(\pm0.1)\times 10^{21}cm^{-2}$, 
$T=6.1(\pm0.1)\times10^5K$, and a luminosity (determined from the normalization constant)
$L=4.9(\pm0.6)\times 10^{36}(D/10~kpc)^2erg\,s^{-1}$.
The observed $N_{H}$ is consistent with the average interstellar absorption towards the source 
\cite[$1.34\times 10^{21}cm^{-2}$,][]{LAB}.

The spectrum during the high-flux state (bottom panel in Figure~\ref{fig:spec}) 
is well fit with the same ONe WD atmosphere model, with  
$N_{\rm H}= 1.5(\pm0.2)\times 10^{21}cm^{-2}$, $T=6.1(\pm0.1)\times10^5$K, 
and a normalization constant a factor $\sim8$ larger, 
corresponding to a luminosity
$L=3.9(\pm0.8)\times10^{37}(D/10~kpc)^2\,erg\,s^{-1}$.

\section{Discussion}

V5116 Sgr was detected as a bright SSS by XMM-Newton 20 months after the nova outburst.
The temperature of the WD atmosphere indicates that residual 
hydrogen burning is occurring in the WD envelope. 
To further confirm the origin of the emission and determine the size of the emitting surface, 
we need an estimation of the distance.
Fortunately enough, the time and magnitude at maximum are quite well determined for 
V5116~Sgr, with a pre-maximum observation only 24~hours before the maximum. 
\cite{lil05} reported $m_V=7.15$ at maximum, and $t_{2}$ (time required 
to decline two magnitudes from maximum) was 6.5$\pm$1.0~days \citep{dob07}.
Using the empirical relation for novae between $M_V$ and $t_2$ \citep{del95} the 
observed $t_2$ implies an absolute magnitude at maximum $M_V=-8.8(\pm0.4)$
(we add a 5\% error to account for the scatter in the $M_V-t_2$ relation).
The photometry at maximum indicates $B-V=+0.48$ \citep{gil05}, and for novae 
at maximum, intrinsic $B-V=0.23\pm0.06$ \citep{ber87}. This implies 
$A_{V}=3.1\,E_{B-V}=0.8\pm0.2$. 
With all this, we estimate the distance to V5116~Sgr to be $d=11\pm3\,kpc$. 
For simplicity, we scale the distance dependent parameters to 10~kpc.
It is worth noticing that the $N_{\rm H}$ obtained from our fits indicates $A_{V}=0.7$ 
\cite[$N_{\rm H}=5.9\times10^{21}E_{B-V}\,\mbox{cm}^{-2}$,][]{zom07}, 
consistent with the value obtained from the observed colors. 

The luminosity during the high flux periods, together with the atmosphere temperature, 
indicate a radius of the emitting object 
$R=6.2(\pm0.9)\times10^8(D/10kpc)\,cm$. This corresponds to 
the whole WD surface emitting during the high flux periods and supports residual 
H-burning as the origin of the SSS emission, rather than any feature related to the 
accretion stream (boundary layer, hot spot, or hot accretion poles).

The most remarkable feature in V5116~Sgr is the SSS light-curve. 
Although the observation lasted just a bit longer than an orbital period, 
we obtain a period compatible with the orbital period
of 2.97~hours found by \cite{dob07}. 
The fact that the temperature is the same for both the high and the low flux phases
indicates that the change in flux cannot be caused by an intrinsic change 
in the H-burning envelope. The luminosity 
during the high flux periods also indicates that the whole WD 
is visible in this state. The decrease by a factor $\sim8$ 
in the flux could be caused by a partial eclipse of the WD. 
The long duration of the low flux phase and short duration of the high flux phase cannot 
be reconciled with a partial eclipse by the secondary star. Alternatively, 
some asymmetric accretion disk could do the job, being responsible for a partial eclipse 
during most of the orbital period, with a 3~ks window during which the whole WD would be visible. 
Since no change is observed in the hydrogen absorption column and it is compatible 
with the average interstellar value both in the low and the high flux spectra, 
the portion of accretion disk producing the partial eclipse should be optically thick 
to the soft X-rays. In the particular case of V5116 Sgr, we have 
checked that an absorbing column of $N_{H}>2\times10^{22}\,cm^{-2}$ would make it
completely opaque to the observed SSS emission.
Longer X-ray observations covering at least two periods 
together with a good timing with the OM camera would be required 
to disentangle the origin of the X-ray light curve behaviour.

\acknowledgments
We thank Jim MacDonald for kindly providing his WD atmosphere 
models. This work is based on observations obtained with XMM-Newton, an ESA science mission with 
instruments and contributions directly funded by ESA Member States and NASA, 
and supported by BMWI/DLR (FKZ 50 OX 0001) and the Max-Planck Society.
This research has been funded by the grants
AYA2004-06290-C02-01 from MEC, 2005-SGR00378 from AGAUR, and by FEDER
funds. GS is supported through DLR (FKZ 50 OR 0405) and CF through an
AGAUR FI fellowship from the Generalitat de Catalunya.

\end{document}